\documentclass[namedreferences]{SolarPhysics}
\usepackage[optionalrh]{spr-sola-addons} 
\usepackage{graphicx}        
\usepackage{amssymb}        
\usepackage{color}           
\usepackage{url}             

\begin{document}

\begin{article}

\begin{opening}

\title{Evolution of the Solar Flare Energetic Electrons in the Inhomogeneous Inner Heliosphere}

\author{Hamish A. S. ~\surname{Reid}$^{1}$\sep
        Eduard P.~\surname{Kontar}$^{2}$
       }
\runningauthor{Reid and Kontar}
\runningtitle{Energetic Electrons in the Inhomogeneous Inner Heliosphere}

   \institute{$^{1}$ LESIA, Observatoire de Paris, CNRS, UPMC, Universit\'{e} Paris-Diderot, 5 place Jules Janssen, 92195 Meudon Cedex, France
                     email: \url{hamish.reid@obspm.fr} \\
              $^{2}$ Shool of Physics and Astronomy, University of Glasgow, G12 8QQ, UK
                     email: \url{eduard@astro.gla.ac.uk} \\
             }

\begin{abstract}
Solar flare accelerated electrons escaping into the interplanetary space and seen as type III solar radio bursts are often detected near the Earth. Using numerical simulations we consider the evolution of energetic electron spectrum in the inner heliosphere and near the Earth. The role of Langmuir wave generation, heliospheric plasma density fluctuations, and expansion of magnetic field lines on the electron peak flux and fluence spectra is studied to predict the electron properties as could be observed by \emph{Solar Orbiter} and \emph{Solar Probe Plus}. Considering various energy loss mechanisms we show that the substantial part of the initial energetic electron energy is lost via wave-plasma processes due to plasma inhomogeneity. For the parameters adopted, the results show that the electron spectra changes mostly at the distances before $\sim20~R_\odot$. Further into the heliosphere, the electron flux spectra of electrons forms a broken power-law relatively similar to what is observed at 1~AU.
\end{abstract}
\end{opening}

\section{Introduction}
Solar flares can efficiently accelerate a large number of electrons to sub-relativistic energies. These energetic electrons
are prominently observed via their hard X-ray emission (see e.g. \opencite{Holman_etal2011}; \opencite{Kontar_etal2011} as recent reviews),
but can also escape into interplanetary space. Energetic electrons that escape flaring regions can be detected
\emph{in-situ} by near-Earth particle detectors \cite{Lin2011}.

Detection of escaping energetic electrons closer to the Sun is via their characteristic emission of both solar
and interplanetary type III radio bursts.  As the electrons travel further away from the Sun the faster electrons
overtake the slower ones creating a positive gradient in velocity space.  When the number of energetic electrons
is sufficiently high Langmuir waves can be generated. The presence of a high level of Langmuir waves can be viewed
via the associated plasma emission, seen as type III radio bursts.

For the electrons exciting type III bursts the transport is governed chiefly by beam-plasma interactions
involving electron-Langmuir wave interactions \cite{ZheleznyakovZaitsev1970,Li_etal1981,Melrose1990}.
This process is fast and the characteristic length of interaction or Langmuir wave generation
by electrons with velocity $v$ and electron density $n(v)$ scales as $\sim vn_p/[\omega_{pe}n(v)]$,
where $\omega _{pe}$ is the electron plasma frequency and $n_{p}$ is the plasma density.
The complicating aspect of the electron transport and Langmuir wave interaction is that Langmuir
waves are effectively scattered and refracted by plasma density fluctuations. This results in a fast change
of Langmuir wave spectrum, which in turn affects the overall evolution of the electron stream travelling
from the Sun to the Earth. Historically, simulations have been used to model the processes of electron beam generation of Langmuir
waves (e.g. \opencite{TakakuraShibahashi1976}) as well the role of density
inhomogeneities (e.g. \opencite{NishikawaRiutov1976}; \opencite{GoldmanDubois1982}; \opencite{Kontar2001b}; \opencite{Li_etal2006}; \opencite{ReidKontar2010}; \opencite{Ziebell_etal2011}).

Recently \inlinecite{ReidKontar2010}, \inlinecite{KontarReid2009} have shown that because of the inhomogeneous plasma of the solar wind,
the spectral break in the peak flux and fluence appears as the electrons travel from the Sun to the Earth.
As a result, the electron flux spectrum (peak and fluence) at 1~AU appears close to a broken power-law with a typical spectral
index below the break around $-2$. Importantly, the exact value is dependent on the level of density fluctuations,
so that the spectral index of electrons below the break is higher for a higher level of density fluctuations.
Above the break the electron spectrum is weakly affected by the Langmuir waves, so can be considered
as `scatter-free' transport. However, these electrons are scatter-free only in respect to Langmuir waves
and could be affected by other plasma waves  (e.g. \opencite{VocksMann2009}; \opencite{BianKontar2010}; \opencite{Threlfall_etal2011}, \opencite{BianKontar2011}, \opencite{Tan_etal2011}).
The simulations by \inlinecite{ReidKontar2010} suggest that the suppression of Langmuir waves changes the electron beam transport.
However, the detailed evolution of the electron spectra from the Sun to the Earth has not been performed.
The question of electron transport and associated Langmuir waves in the inner heliosphere
becomes particularly important in the view of anticipated observations by ESA's \emph{Solar Orbiter} and NASA's \emph{Solar Probe Plus}.

In this paper, using a time-dependent injection,  we investigate the detailed evolution of energetic electrons
in the inner heliosphere. Specifically, we investigate the evolution of the spectral index change
as the function of distance for both the electron fluence and the electron peak flux, the energy
loss rate as a function of distance and the role of electron injection time during a solar flare.

\section{Electron transport of deka-keV electrons: model}
\label{ref:Sec2}

To model the transport of an electron beam from the Sun to the Earth we have taken a similar
approach to the work that was done in \inlinecite{ReidKontar2010}.  We use a set of two kinetic equations describing the motion
of electrons along the radially expanding magnetic field lines and their interaction with plasma via emission
and absorption of Langmuir waves. Whilst the production of Langmuir waves is dealt with in a self-consistent manner, the non-linear effects of Langmuir waves scattering off ions and Langmuir wave coupling with ion sound waves is not considered (e.g. \opencite{KontarPecseli2002}; \opencite{Ziebell_etal2011}).  Both the electrons and Langmuir waves exchange energy
through the resonant interaction $\omega_{pe}=kv$ described by quasilinear terms [first term on the right hand sides \cite{Drummond_Pines1962,Vedenov_etal1962}]
\begin{eqnarray}
\frac{\partial f}{\partial t} + \frac{v}{(r+r_0)^2}\frac{\partial}{\partial r}(r + r_0)^2f =
\frac{4\pi ^2e^2}{m^2}\frac{\partial }{\partial v}\frac{W}{v}\frac{\partial f}{\partial v}  \;\cr +\frac{4\pi n_e e^4}{m_e^2}\ln\Lambda\frac{\partial}{\partial v}\frac{f}{v^2},
\label{eqk1}
\end{eqnarray}
\begin{eqnarray}
\frac{\partial W}{\partial t} + \frac{\partial \omega_L}{\partial k}\frac{\partial W}{\partial r}
-\frac{\partial \omega _{pe}}{\partial r}\frac{\partial W}{\partial k}
= \frac{\pi \omega_{pe}(r)}{n_e}v^2W\frac{\partial f}{\partial v} \;\;\;\;\;\;\;\cr - (\gamma_{c} +\gamma_L )W + e^2\omega_{pe}(r) v f \ln{\frac{v}{v_{Te}}},
\label{eqk2}
\end{eqnarray}
where  $f(v,r,t)$ is the electron beam distribution function, and $W(v,r,t)$ the spectral energy density
of Langmuir waves.
The second term on the left had side of Equation (\ref{eqk1}) describes the propagation of electrons along the radial direction $r$.
It also models a radial decrease in density of the electron beam as the beam fills the radial expanding magnetic field.
This expansion is determined by the constant $r_0=3.4\times10^9$~cm which is chosen to model a cone of expansion
with an angle of $33.6^o$.  The two terms on the right hand side of Equation (\ref{eqk1}) model the quasilinear diffusion of electrons
in the presence of Langmuir waves and the collisional interaction of electrons off ions in the background plasma
respectively.

The second term on the left hand side of Equation (\ref{eqk2}) describes the propagation of Langmuir waves along the radial
direction $r$, where $\partial \omega_L/\partial k$ is the group velocity of Langmuir waves.  The third term on the left hand side
of Equation (\ref{eqk2}) describes the refraction of Langmuir waves at the density gradients in the background heliospheric plasma.
The first term on the right hand side of Equation (\ref{eqk2}) describes the growth rate of Langmuir waves from an unstable electron
beam through resonant interaction.  $\gamma_c$ and $\gamma_L$ correspond to the collisional absorption of Langmuir waves
and Landau damping respectively, where $\gamma_{c} = {\pi n_e e^4}\ln\Lambda/(m_e^2 v_{Te}^3)$ and $\gamma_L=2\sqrt{\pi}\omega_{pe}(r)\left(v/v_{Te}\right)^3\exp\left(-{v^2}/{v_{Te}^2}\right)$.
The last term on the right hand side of Equation (\ref{eqk2}) represents the spontaneous wave generation \cite{ZheleznyakovZaitsev1970,TakakuraShibahashi1976,Hannah_etal2009}.  

\subsection{Density model in the solar corona and the inner heliosphere}

To model non-thermal electron transport through the heliosphere, taking into account self-induced Langmuir waves,
an accurate model of the background electron density is required. Previous work \cite{ReidKontar2010} investigated the dependency
of Langmuir wave generation from the Sun to the Earth with background density fluctuations. Such fluctuations are able to suppress
Langmuir wave growth and as such the turbulent nature of the solar wind must be taken into account for accurate modelling of electron
beam propagation.  Following the same approach as \inlinecite{ReidKontar2010}, we initially model the large scale decrease in electron
density in the solar wind $n_0(r)$ using the equations for a stationary spherical symmetric solution \cite{Parker1958}
with normalisation factor by \inlinecite{Mann_etal1999}.  
\begin{equation}\label{sol1}
r^2n_0(r)v(r)= C= const
\end{equation}
\begin{equation}\label{sol2}
  \frac{v(r)^2}{v_c^2}-\mbox{ln}\left(\frac{v(r)^2}{v_c^2}\right)=
  4\mbox{ln}\left(\frac{r}{r_c}\right)+4\frac{r_c}{r}-3
\end{equation}
where $v_c\equiv v(r_c)=(k_BT_e/\tilde{\mu}m_p)^{1/2}$, $r_c=GM_s/2v_c^2$, $T_e$ is the electron temperature, $M_s$ is the mass of the Sun, $m_p$ is the proton mass and $\tilde{\mu}$ is the mean molecular weight. The constant appearing above is fixed by satellite measurements near the Earth's orbit (at $r = 1$~AU, $n =6.59$~cm$^{-3}$) and equates to $6.3\times 10^{34}$ s$^{-1}$.  The density $n_0(r)$ is calculated by numerical integration of Equations (\ref{sol1}) and (\ref{sol2}).  The density profile  $n(r)$ is then modelled by adding density fluctuations whose power density spectrum takes the form of a power-law with spectral index $-5/3$,
\begin{equation}\label{pertm1}
n(r) = n_0(r)\left[1 + C \sum_{n=1}^{N}  \lambda_n^{\beta/2} sin( 2\pi r/\lambda_n + \phi_n)\right],
\end{equation}
where $\lambda_n$ is the wavelength of the perturbations with random phase $\phi_n$, $\beta=5/3$.
$C$ is a constant which defines the r.m.s. level of density fluctuations $\sqrt{{\langle \Delta n(r)^2 \rangle}/{\langle n(r) \rangle^2}}$
or $\Delta n(r)/n(r)$ for short.  The range of wavelengths modelled is $10^7\leq \lambda_n \leq 10^{10}$~cm.
We found previously \cite{ReidKontar2010} that a high level of density fluctuations ($10\%$),  an observed value near the Earth \cite{Celnikier_etal1987},
all the way from the Sun to the Earth contradicts to the high level of Langmuir waves required to explain the observed solar type III radio bursts.
The relative level of density fluctuations $\Delta n(r)/n(r)$ is assumed to be decreasing towards the Sun 
from $10\%$ at 1~AU using
\begin{equation}
\frac{\Delta n(r)}{n(r)} = \left(\frac{n_0(1AU)}{n_0(r)}\right)^\psi \frac{\Delta n(1AU)}{n(1AU)}
\end{equation}
where $\psi$ determines the rate at which levels of density fluctuations rise from the Sun to the Earth.
It was found through a comparison with observations that a reasonable value for $\psi=0.25$.

\section{Initial electron injection} \label{ref:Sec3}

To model the injection of energetic electrons into the solar corona,
we use an electron distribution function $f(v,r,t)$, which varies separately in velocity, space and time
\begin{equation}\label{eq:init_f}
f(v,r,t) = g_0(v)h_0(r)i_0(t).
\end{equation}
The injected electron distribution function varies with velocity as a single power-law,
consistent with HXR observations.
\begin{equation}
g_o(v) = \frac{n_{b}(2\delta -1) }{v_{\rm{min}}}\left(\frac{v_{\rm{min}}}{v}\right)^{2\delta}
\end{equation}
where $\delta$ is the spectral index of the electron beam (in energy space), and $n_b$ is time integrated beam density taken
as $10^7~\rm{cm}^{-3}$, such that realistic peak flux and fluence values are measured near the Earth \cite{Lin1985,Krucker_etal2007,Krucker_etal2009}.
$v_{\rm{min}}$ is the minimum beam velocity which is set at $3~v_{Te}$, where $v_{Te}=7.8\times10^8~\rm{cm~s}^{-1}$
is the background thermal electron velocity associated with a 2~MK plasma. $v_{\rm{min}}$ is defined close enough
to $v_{Te}$ such that Langmuir waves are suppressed via Landau damping.  The highest velocity modelled is $2\times10^{10}~\rm{cm~s}^{-1}$,
which is equivalent to $115$~keV under the non-relativistic assumption.  High velocities would require consideration of relativistic dynamics
and are not required considering the range of energies usually involved in impulsive solar electron beam wave-particle interaction.

The energetic electrons are injected in a finite volume near $r=0$, which corresponds to height $h=5\times10^9$~cm (50 Mm) at a frequency of 415 MHz.  The electron distribution takes the form of a Gaussian
\begin{equation}
h_0(r)=\exp\left(\frac{-r^2}{d^2}\right),
\end{equation}
where $d$ is the characteristic length scale of the electron acceleration region or one dimensional electron injection region.
$d$ is set to $10^9$~cm (10~Mm) which is the longitudinal scale of the electron acceleration region found
in the analysis by \inlinecite{Reid_etal2011}. The length of the simulation box is set to $1.2$~AU which is a typical
distance for electrons to reach the Earth, being larger than 1~AU due to the curvature of the assumed Parker spiral.

Temporally, the electron distribution function also varies as a Gaussian with two characteristic timescales
\begin{equation}
i_o(t)=\frac{1}{0.5\sqrt{\pi}(\tau_1+\tau_2)}\exp\left(\frac{-(t-t_0)^2}{\tau^2}\right),
\label{eqn:time_inj}
\end{equation}
where $\tau=\tau_1$ in the rise, for time $t<t_0$ and  $\tau=\tau_2$ in the decay time $t\geq t_0$.  $\tau_1$ and $\tau_2$ represent the characteristic rise
and decay times with $\tau_1 < \tau_2$.  By assuming a common acceleration region for both upward
and downward propagating electron beams, observational values for $\tau_1$ and $\tau_2$ can be obtained from hard X-ray measurements
of solar flares \cite{Holman_etal2011,Kontar_etal2011}.  We use $\tau_1=10$~s and $\tau_2=45$~s which are the typical values for large flare hard X-ray rise and decay times.  The value of $t_0$ is set at $t_0=4\tau_1$ to allow sufficient time for the rise of electron injection.
The approximation of a Gaussian rise time will not capture the fine structure of electron acceleration in the corona, which is often
evident in both X-ray and type III radio observations.  However, it adequately reproduces the overall
temporal behaviour of the electron acceleration.

Observationally a correlation in the spectral index has been found for prompt electron events at 1~AU
and HXR emitting electrons with energies $>50$~keV \cite{Krucker_etal2007}.  Such energies are likely
to travel to 1~AU without altering the spectrum by generating Langmuir waves \cite{KontarReid2009,ReidKontar2010}.
Using the same approximation of a common acceleration region, we set the injected spectral index $\delta$ to be similar values
found from hard X-ray observations.  Although we note that a typical X-ray spectral indices show a so-called `soft-hard-soft' behaviour
in time (e.g. \opencite{ParksWinckler1969}; \opencite{Benz1977}; \opencite{Holman_etal2011}), where $\delta$ starts high (soft spectrum), then goes low during
the most intense period of emission (hard spectrum) and the returns high
afterwards (soft spectrum).  We vary $\delta$ between $8$ and $4$ with the same time dependence
used for electron injection (Equation (\ref{eqn:time_inj})).

We assume an initial spectral energy density of Langmuir waves at the thermal level
\begin{equation}\label{init_w}
W(v,r,t=0) = \frac{k_BT_e}{4\pi^2}\frac{\omega_{pe}(r)^2}{v^2}\log\left(\frac{v}{v_{Te}}\right).
\end{equation}
The thermal level of Langmuir waves represents the spontaneous emission of Langmuir waves from the background Maxwellian plasma.  The spontaneous generation of Langmuir waves (the last term on the right hand side of Equation (\ref{eqk2})) only deals with the generation of Langmuir waves from the flare accelerated electron beam.  Therefore to model the spontaneous emission of Langmuir waves at times $t>0$ we set the minimum level of Langmuir waves to be this thermal equilibrium $W(v,r,t=0)$.

\section{Energetic particle energy loss in the corona and in the heliosphere}

It has been found observationally that the peak flux and fluence of impulsive electron beams can take the form of a broken power-law at the Earth \cite{Wang_etal1971,Lin1974}.  The break takes the form of a knee with lower energies having a smaller spectral index. A recent statistical study \cite{Krucker_etal2009} using the 3-D Plasma and Energetic Particle instrument \cite{Lin_etal1995} on the WIND spacecraft was carried out on 62 impulsive events.  The average break energy was found to be around 60~keV although breaks were detected as low as 30~keV and sometimes above 100~keV.  Weaker events are also detected only at low and high energies depending upon the background flux of electrons.

Recent work \cite{KontarReid2009,ReidKontar2010} explains that a broken power-law spectrum at the Earth is created from a single power-law injected electron spectrum if the generation and absorption of Langmuir waves are taken into account. Specifically the broken power-law forms as a combined process between the electron beam inducing Langmuir waves and the density inhomogeneity changing the phase velocity of the Langmuir waves.  Above the break energy the electron beam is too dilute to generate Langmuir waves in the background plasma.  However, below the break, Langmuir wave interaction with the electrons flattens the electron distribution function in velocity space. As the result, the peak flux and fluence spectrum flattens with the density inhomogeneity controlling the flattening.

\begin{figure} \center
\includegraphics[width=0.99\columnwidth]{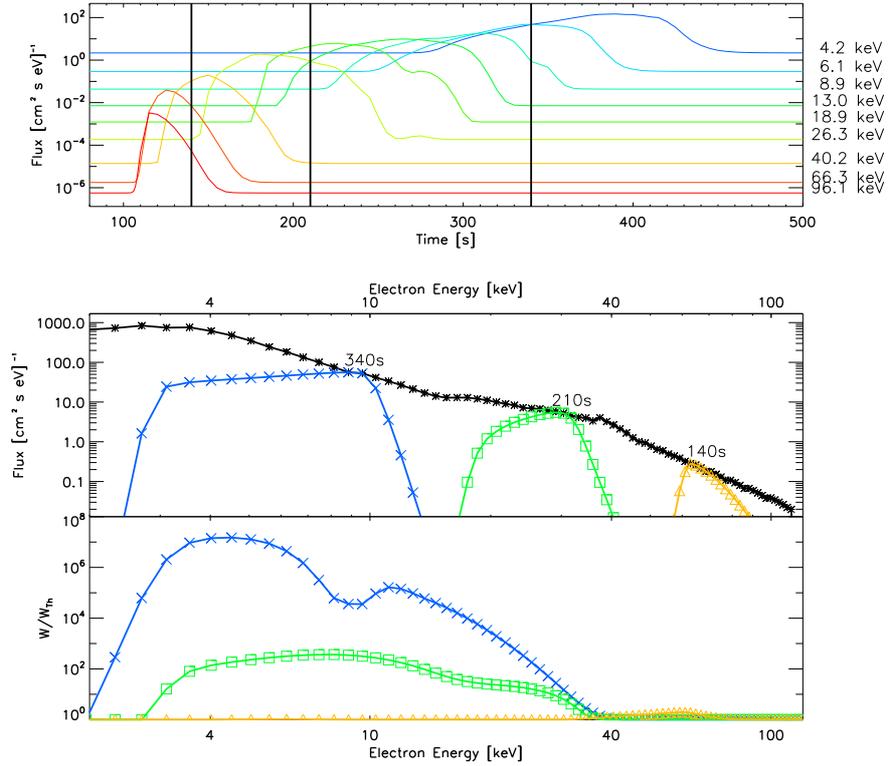}
\caption{The electron distribution function and Langmuir wave spectral energy density at 21.5 solar radii from the injection point.  Top: the electron beam flux as a function of time with energy channels denoted on the right hand side.  The three black lines represent the times 140 s, 210 s, 340 s used for the other two graphs.  An artificial power-law background flux has been added for clarity.  Middle: the electron beam flux as a function of energy at three points in time.  The black curve represents the peak flux spectrum.  Bottom: the spectral energy density of Langmuir waves normalised by the initial thermal level $W_{Th}$ at the same points in time as the electron flux.  Colours and symbols in the middle and bottom graph represent the same times.  Note the lower energy electrons arriving at 21.5 solar radii at later times are associated with larger wave
amplitudes.}
\label{fig:maxfpos}
\end{figure}

An example of this behaviour is shown in Figure \ref{fig:maxfpos} at a distance of 21.6~$R_\odot$ for a simulation using the equations from Section \ref{ref:Sec2} and initial parameters from Section \ref{ref:Sec3}.  The top panel shows the lightcurves of the simulations (electron flux as a function of time) using the same energy bins as the WIND 3DP instrument.  A power-law artificial background has also been added.  One can observe higher energy particles arriving before low energy particles.  The middle and bottom panels show the electron flux and normalised spectral energy density as a function of electron energy at the different times.   When the high energy particles ($> 40$ keV) arrive at 21.6~$R_\odot$ we can observe no significant Langmuir wave growth.  Consequently the high energy electron distribution is not changed through quasi-linear interaction.  Lower energy particles ($<40$~keV) have a higher flux and therefore are susceptible to wave-particle interactions.  Note the broadened flux distribution in energy space at time $t=340$~s and the corresponding high level of Langmuir waves above the thermal background.  The peak flux spectrum (black line in the middle graph) has noticeably flattened below 40 keV.

\subsection{Collisional energy losses}


The collisional energy losses for the electron beam are modelled by the last term in Equation (\ref{eqk1}).  Electron-electron collisions are important for low energy electrons in a dense background plasma of the corona
near the injection site.  As such, the electron beam only suffers significant energy loss from collisions near the acceleration site.
Exactly how close to the acceleration site collisions play an important role is required for understanding the electron beam energetics.

\begin{figure} \center
\includegraphics[width=0.8\columnwidth]{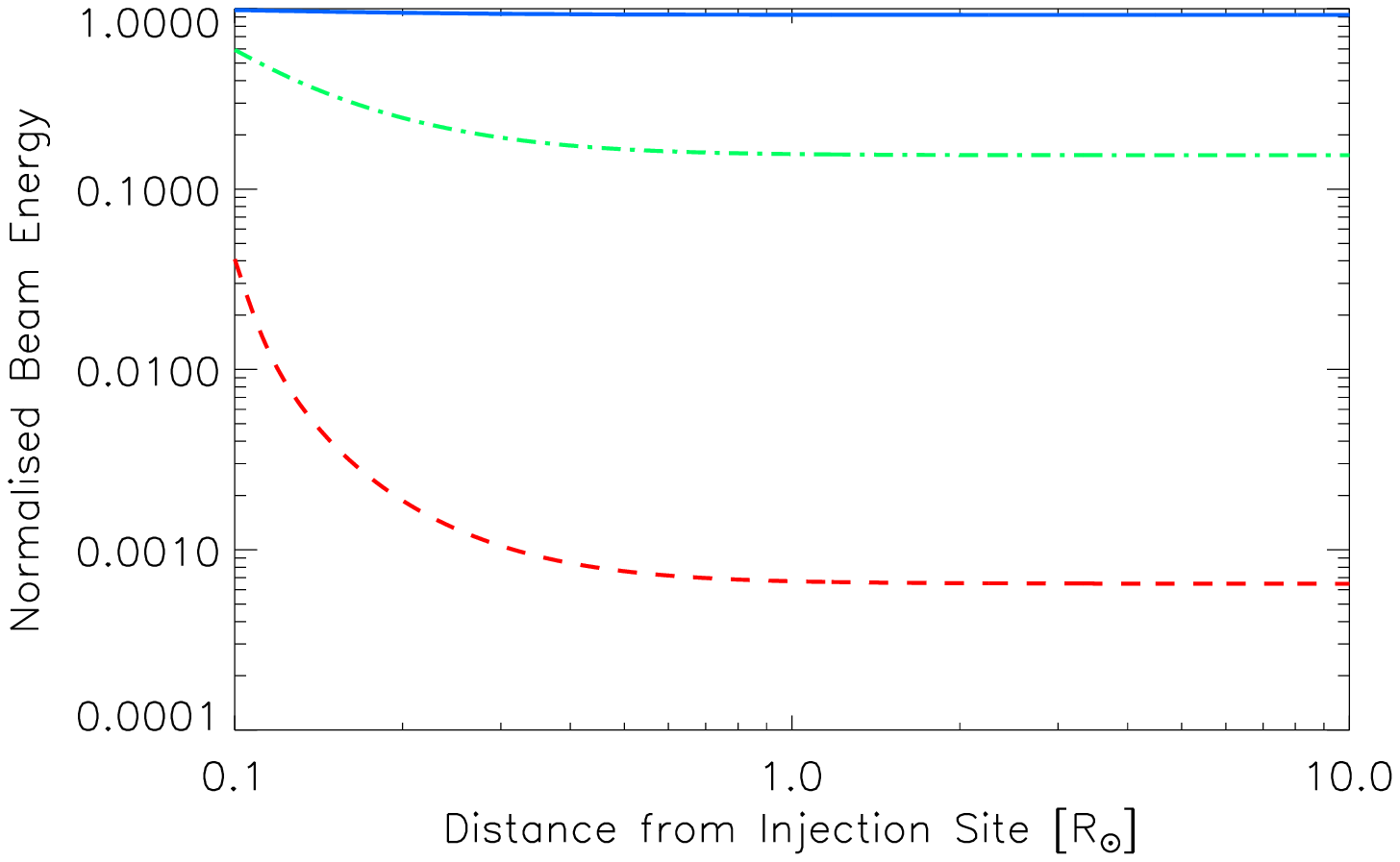}
\caption{The electron beam energy as a function of distance from the acceleration site when electron-electron collisions are taken into account.  The beam energy is normalised by the electron beam when no collisions are present.  The red dashed line corresponds to the energy range $E_{\rm{min}} \leq E \leq 6$~keV.  The green dot dashed line corresponds to the energy range $6~\rm{keV} \leq E \leq 40$~keV.  The blue solid line corresponds to the energy range $40~\rm{keV} \leq E \leq 100$~keV.}
\label{fig:cp}
\end{figure}

Figure \ref{fig:cp} shows the ratio of electron beam energy with and without collisions for different energy bands as a function of distance
from the acceleration site.  We can see that collisions are really only important at small distances $r \leq 1~\rm{R_\odot}$ from the acceleration site.
Moreover, the majority of energy losses occurs with low energy electrons $\lesssim 6$~keV.  Electrons with energy $E \gtrsim 40$~keV are largely unaffected
by collisions.  The resultant electron beam leaves the corona with approximately $0.7\%$ of its injected energy.  This percentage is heavily dependent upon the initial conditions of the electron beam, for example the starting height of the electron beam or the spectral index of the electron beam.

\subsection{Generation of Langmuir waves in uniform plasma} \label{uniform}

To understand the energy loss an electron beam undergoes travelling through the turbulent medium of the solar wind we must first consider a uniform plasma.  It was initially postulated \cite{Sturrock1964} that an electron beam would lose all its energy to Langmuir waves over a very short distance.  Later it was found through numerical simulations \cite{TakakuraShibahashi1976,MagelssenSmith1977} that an electron beam was able to propagate long distances by forming a beam-plasma structure.  The structure travels through the heliosphere with the characteristic velocity of the energetic electrons.  Langmuir waves, whose group velocity is orders of magnitude lower than the electron's kinetic velocity, are generated at the front of the electron beam and absorbed at the back of the electron beam.  The generation of a beam-plasma structure is a result of the finite spatial width of an electron beam.  The converse, a spatially uniform beam, would not undergo the bump-in-tail instability assuming a power-law electron distribution and the absence of collisions, as fast electrons would no longer be able to outpace slower electrons.

It was mentioned as early as \inlinecite{TakakuraShibahashi1976} that there is very little energy loss when an electron beam propagates in a beam-plasma structure.  Indeed, the quasilinear terms of Equations (\ref{eqk1}) and (\ref{eqk2}) conserve energy, momentum and density.  Energy is temporarily exchanged between the particles and the waves.  In the presence of a uniform background plasma all the energy that is transferred to Langmuir waves at the front of the beam should be re-absorbed by the electrons at the back of the beam.  By running a simulation which just considered the quasilinear interaction of the electron beam we compared the energetics to a free-streaming simulation.  There was no significant net loss of energy when waves are produced (under the assumption of no binary Coulomb collisions) for energies above where Landau damping is significant.  However, the inner heliosphere is highly anisotropic so this simple situation where beam energy is conserved does not apply to type III producing electron beams.

\subsection{Generation of Langmuir waves and collisional damping of waves}

One process which can reduce the energy of a beam-plasma structure is collisional damping of Langmuir waves in the background plasma where the damping rate is $\gamma_{c} \approx \frac{\pi n_e e^4}{m_e^2 v_{Te}^3}\ln{\Lambda}$.  Similar to particle collisions, the collisional damping of waves is heavily dependent upon the background density of electrons.  As a result it has the greatest effect close to the Sun.  However, an important difference is that collisional damping of waves is independent on the phase velocity of Langmuir waves.

The energy contained in Langmuir waves is much smaller than the kinetic energy of the electrons.  Therefore, we do not expect wave collisions to reduce the total energy in the beam-plasma structure to a similar extent as particle-particle collisions.  Given the initial conditions described in Section \ref{ref:Sec3} we find that the effect of wave collisions is negligible on the total energy.  Wave-particle instability in the beam of particles does not occur instantly but takes around 2~$R_\odot$ of propagation to be noticeable (see Section \ref{ref:Sec7} for a discussion).  The background density at such distances is too small for significant collisional absorption of Langmuir wave energy.  The collisional damping of waves will play an important role deep in the corona.  However, it is not the dominant energy loss process for electrons travelling through the heliosphere.

\subsection{Energy losses due to inhomogeneous plasma}

The spatial gradient of the background electron density plays an important role in the dynamics of Langmuir waves.  It has been shown before (e.g. \opencite{Kontar2001}) that Langmuir waves are refracted to different k-vectors by the inhomogeneous background electron density gradient (Term 2 of the left hand side of Equation (\ref{eqk2})).  The radial dependence in this process is mostly governed by the characteristic scale of plasma inhomogeneity $L(r) = 2 n_e(r) (\partial n_e(r)/\partial r)^{-1}$.  For a simple solar wind plasma model where density only decreases $L(r)$ is strictly negative.  When fluctuations are added to mimic the turbulent nature of the solar wind $L(r)$ not only becomes positive in parts but varies more in magnitude.  Such fluctuations in the background electron density can suppress Langmuir wave growth by moving Langmuir waves to higher and lower k-vectors, out of resonance with inducing electrons.

\begin{figure} \center
\includegraphics[width=0.8\columnwidth]{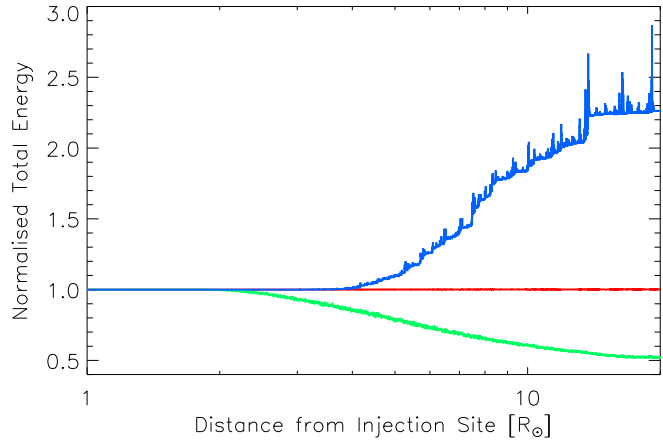}
\caption{The total energy in the beam-plasma structure (waves and particles) as a function of distance when background plasma inhomogeneity is taken into account.  Energy is normalised using the simulation  where Langmuir waves are induced from the quasilinear term but the inhomogeneity term was ignored. The red line corresponds to the energy range $E_{\rm{min}} \leq E < 6$~keV.  The green line corresponds to the energy range $6~\rm{keV} \leq E < 40$~keV.  The blue line corresponds to the energy range $40~\rm{keV} \leq E < 100$~keV.}
\label{fig:nu}
\end{figure}

To examine how the electron beam varies with distance the total beam-plasma energy has been plotted for various energy ranges in Figure \ref{fig:nu}.  Values were normalised using the simulations from Section \ref{uniform} where wave interaction was taken into account but not the role of inhomogeneous plasma.  We define three energy channels to analyse the energetics  $E_{\rm{min}} \leq E < 6$~keV, $6~\rm{keV} \leq E < 40$~keV and $6~\rm{keV} \leq E < 40$~keV.  Resonant Langmuir waves whose phase velocity is the same as the electron velocities within the three energy ranges are included in the energetics.

In Figure \ref{fig:nu} we observe no net change in energy below $6$~keV.  The bulk of energy is stored in electrons near $E_{\rm{min}}$ where Langmuir waves are absorbed due to Landau damping by the background Maxwellian plasma.  Between $6$ and $40$~keV, beam-plasma energy starts to decrease around $2~R_\odot$ when significant levels of Langmuir waves become induced by the electron beam.  By $20~R_\odot$ this energy range has lost approximately half the energy it would have without the effect of inhomogeneous plasma.  Conversely the higher energy channel between $40$ and $100$~keV gains energy as Langmuir waves are absorbed.  This occurs at a greater distance than $2~R_\odot$ because it is linked to positive values of $L(r)$ which moves wave energy to lower values in k-space (higher phase velocity).  It is not until the beam propagates further away from the Sun that effects from the background electron density turbulence are significant over the radially decreasing density decrease (see \opencite{ReidKontar2010} for further details).  When positive values of $L(r)$ become significant, wave energy is moved to higher phase velocities and can be absorbed by higher energy electrons.  Positive values of $L(r)$ can be observed as the spikes in the high energy channel and correspond to waves at high phase velocities.  Thus, inhomogeneity causes a spectral shift in energy.

\subsection{Energy loss from the Sun to the Earth}

\begin{figure} \center
\includegraphics[width=0.8\columnwidth]{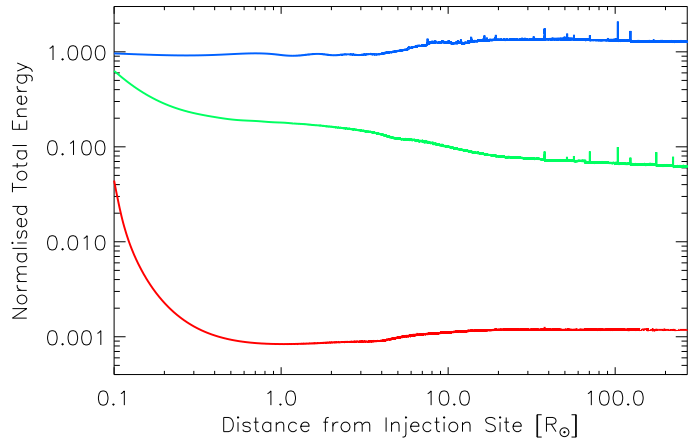}
\caption{Total energy of the beam-plasma structure (waves and particles) as a function of distance from the Sun in three different energy bands.  The energy is normalised using the simulation where electrons are free-streaming only.  The red line corresponds to the energy range $E_{\rm{min}} \leq E < 6$~keV.  The green line corresponds to the energy range $6~\rm{keV} \leq E < 40$~keV.  The blue line corresponds to the energy range $40~\rm{keV} \leq E < 100$~keV.}
\label{fig:all}
\end{figure}

After looking at the energy loss from the individual terms, we now consider propagation from the Sun to the Earth using all physical processes defined previously in Equations (\ref{eqk1}) and (\ref{eqk2}).  Figure \ref{fig:all} shows how the energetics in the previously defined energy channels varies with radial distance from the Sun normalised by the case where only free-streaming electrons are considered.  We observe the same initial high level of energy loss of the electron beam in the corona which was predicted in our analysis of the coulomb collisions.  The dense plasma of the corona damps the low velocity electrons which contain the majority of the total energy in the electron beam.

After $1~R_\odot$ the background plasma rarefies and inhomogeneity plays the dominant role for the electron beam energy loss.  By 1~AU the energy range $6~\rm{keV} \leq E < 40$~keV has lost nearly 1 order of magnitude of energy due to inhomogeneity with respect to the free-streaming case.  Conversely the higher energy range $40~\rm{keV} \leq E < 100$~keV has gained a small amount of energy by absorbing wave energy refracted to higher phase velocity.

\section{Electron spectrum evolution}

\begin{figure} \center
\includegraphics[width=0.99\columnwidth]{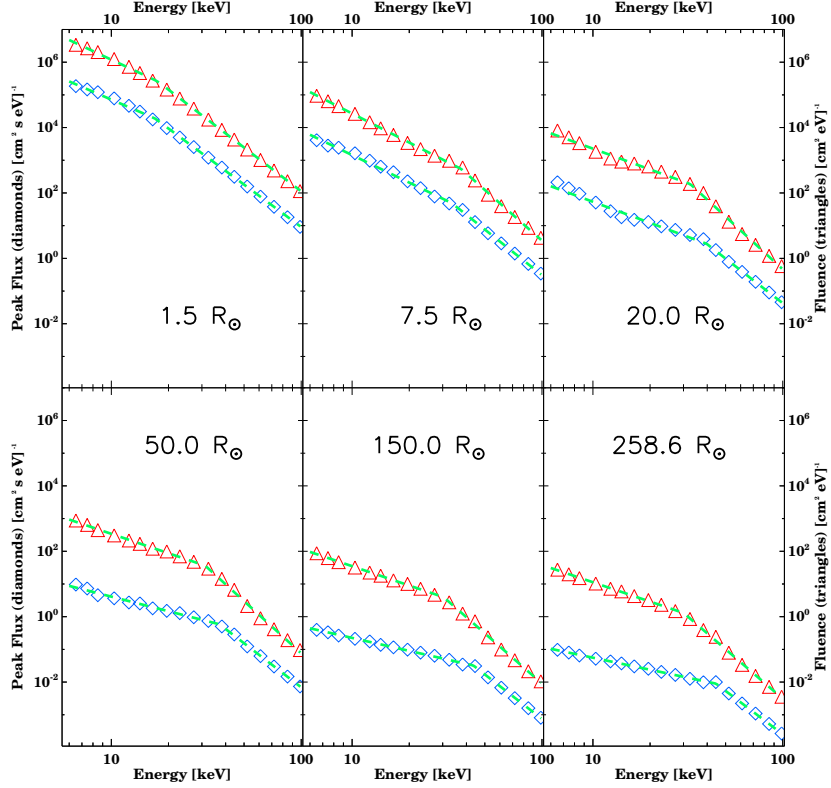}
\caption{Fluence (red triangles) and peak flux (blue diamonds) spectrum of energetic electrons at different distances from the injection site.  The bottom right panel shows the spectrum at 1.2~AU.  A green best fit line has been drawn using a broken power-law function for all curves.}
\label{fig:spectrum_all}
\end{figure}

Although a break in the power-law spectra appears to be formed in the electron spectrum from propagation effects,
it is not obvious how the break forms as a function of distance from the Sun.
 Our previous work \cite{ReidKontar2010} suggests that high energy ($>20$~keV) electrons only participate in wave-particle interactions
 close to the Sun before their flux rarefies due to the expansion of the magnetic field.  In this scenario the power-law break would be formed
 quite early in the propagation of the electron beam towards the Earth.  Any predictions concerning electron beam dynamics
 between the Sun and the Earth is particularly relevant with the upcoming missions of \emph{Solar Orbiter} and \emph{Solar Probe Plus} later this decade.
 As such we investigate how the electron spectrum both in peak flux and fluence (time integrated flux) varies radially
 from the Sun to the Earth in our simulations (Figure \ref{fig:spectrum_all}).

\begin{figure} \center
\includegraphics[width=0.8\columnwidth]{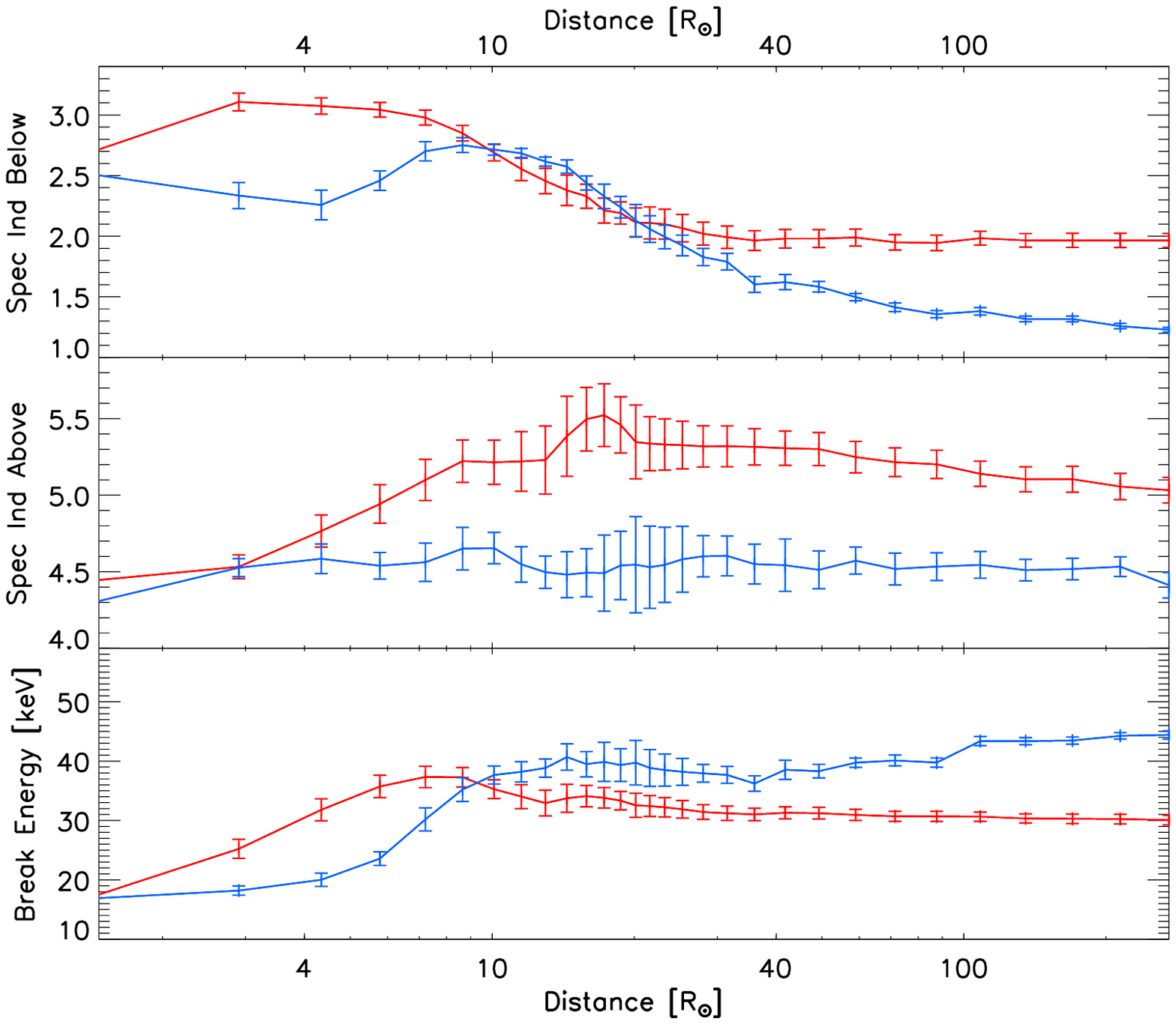}
\caption{The broken power-law fit parameters for the fluence (red) and peak flux (blue) spectra as a function of distance from the Sun.  Top: spectral index below the break energy.  Middle: spectral index above the break energy.  Bottom: break energy.}
\label{fig:spectrum_fit}
\end{figure}

To investigate the evolution of the electron beam spectrum, we fit both the peak flux and fluence with a broken
power-law as a function of distance.  The free parameters for the fit are the break energy, the flux or fluence at the break energy
and the spectral index both below and above the break energy.  The fit is made between 6 and 100 keV.  Figure \ref{fig:spectrum_fit} shows the variation of three free parameters for the fit as a function of distance.  The weighting for the fit was $1/y^2$ such that $\chi^2=\sum{(y-y_{fit})^2/y^2}$ allowing the closest logarithmic fit to the data.  By assuming an implicitly good fit we have
scaled the one sigma errors with the chi-squared values using $\sigma_c=\sigma \sqrt{\chi^2/\rm{DOF}}$,
where DOF represents the degrees of freedom in the fit, discerned by the number of points minus the number of fit parameters.  To analyse Figure \ref{fig:spectrum_fit} we will consider both the peak flux and fluence at specific distances
from the Sun that are plotted in Figure \ref{fig:spectrum_all}.

Initially, as shown in the previous section on energetics, the collisions dominate the transport of the low energy electrons inside $1~R_\odot$.  Such a scenario will not be ubiquitous of all electron beams.  However, due to the high characteristic time of the temporal electron beam injection and the starting background electron density we chose it is expected.  Figure \ref{fig:spectrum_all} shows a snapshot of the peak flux and fluence spectrum at $1.5~R_\odot$.  The data has been logarithmically binned in energy space to show a clear representation. The green dashed line shows a best fit to the data using a broken power-law function.   The break energy is very low ($<20$~keV) on account of the spectral change from Coulomb collisions affecting low energy electrons.  The spectral index above the break energy for both the peak flux and fluence is around 4.5 which is exactly what would be expected for free streaming electrons with initial injection distribution described by Equation (\ref{eq:init_f}).

We have already shown that wave-particle interactions have become energetically important for the electron beam by 2.5~$R_\odot$.  Again, we highlight this distance is defined by the initial injection parameters of the electron beam simulated.  When wave-particle interactions become important the spectral index of the beam starts to change below $40$~keV.  Figure \ref{fig:spectrum_all} shows a snapshot of the spectrum at $7.5~R_\odot$.  A bump in the fluence spectrum can be seen around 40~keV along with spectral flattening below 30 keV.  The peak flux also shows spectral flattening below 40 keV.  The spectral flattening is enough that the reduction in flux from collisions at low energies $<10$~keV is no longer visible.  The bump in fluence spectrum is caused by the absorption of wave energy which has been shifted to higher phase velocities by positive background electron density gradients.  The bump has caused an increase in the fluence spectral index above the break energy from 4.5 to around 5.  The break energy for both the peak flux and fluence has increased to between 30 and 40 keV.

From Figure \ref{fig:spectrum_fit} we observe the spectral index below the break energy decreasing for both peak flux
and fluence spectra till about $20~R_\odot$.  We note that the energy spectrum observed in the simulations
is not an exact broken power-law but more of a triple power-law. However, it can still be approximated with a double power-law fit.
This can be seen from the poorness of the fit represented by the increased errors presented in Figure \ref{fig:spectrum_fit}.
After $20~R_\odot$ the spectral parameters change slowly.   The peak flux and fluence spectra
in Figure \ref{fig:spectrum_all} are very similar at distances of $50$~R$_\odot$ (0.23~AU), $150$~R$_\odot$ (0.7~AU)
and $258.6$~R$_\odot$ (1.2~AU).  The spectral index below the break for the peak flux decreases the most from around 2 to 1.2.
while the fluence spectral index below the break remains almost constant.  The density inhomogeneity also
leads to re-acceleration of beam electrons evident in the bump in the fluence spectrum. This is related
to Langmuir waves shifted to smaller $k$ (larger velocities) by the density inhomogeneities
and consequently reabsorbed by the higher energy electrons, which results in appearance
of accelerated electrons \cite{NishikawaRiutov1976,Kontar2001b}. The bump in the fluence also
makes the break energy in the fluence spectrum slightly ambiguous compared to the sharp break
in the peak flux spectrum.

\section{Electron density turbulence}

\begin{figure} \center
\includegraphics[width=0.7\columnwidth]{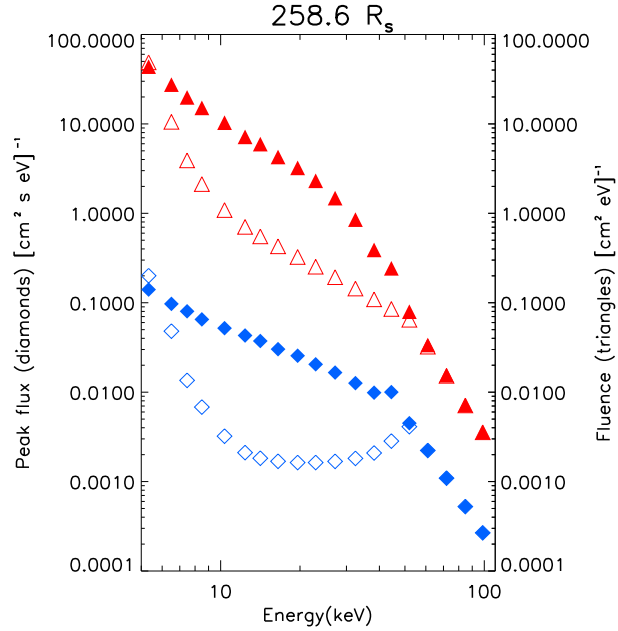}
\caption{The peak flux and fluence of the electron beam at 1.2~AU with density fluctuations (filled symbols) and without fluctuations (empty symbols) in the heliosphere assumed.  Without considering density fluctuations the peak flux has a flat distribution between 10 and 40 keV that does not agree with observations near the Earth.}
\label{fig:noinhomo}
\end{figure}

The level of density fluctuations $\Delta n(r)/n(r)$ directly affects the amount of Langmuir waves produced by the electron beam
as it travels through the inner heliosphere.  If no density fluctuations are present a higher level of Langmuir waves is produced.
To highlight their importance we ran a simulation which only modelled a radially decreasing density and compared the spectrum
at the Earth to a simulation which included density fluctuations.  The corresponding peak flux and fluence spectrum
is presented in Figure \ref{fig:noinhomo}, where the filled and empty symbols represent the case with
and without density fluctuations respectively. The increased level of Langmuir waves from a lack of density fluctuations
caused increased flattening of the spectrum below the break energy in both fluence and peak flux.
Indeed the flattening is so extreme that the peak flux spectrum is a plateau (spectral index around 0) between the energies 10-40 keV.
Such a situation is never observed in-situ with spacecraft.

The cause of such a flat spectrum is the presence of an anomalously large magnitude of Langmuir waves
when the electrons below 40 keV arrive at the Earth.  A combination between a plateau forming in the electron
distribution function and wave refraction causes this constant maximum between 10 and 40 keV.
The observed increase at 10 keV is caused by Landau damping absorbing Langmuir waves at low energies.
The electron distribution function consequently cannot spread as much in energy space and, as number density is conserved, we observe both as increase in peak flux and fluence.  Whilst this scenario of zero density fluctuations is unrealistic for the solar wind,
a relatively low level of density fluctuations could be responsible for observations of solar beam-plasma structures
associated with type III radio bursts at distances as large as 4.3~AU \cite{Buttighoffer_etal1995}.

\section{Discussion and conclusions} \label{ref:Sec7}

We have investigated the changing energetics of a solar electron beam as it travels through the inner heliosphere.
The majority of energy loss for an electron beam is shown to happen in the dense solar corona and comes from Coulomb
collisions of electrons with ions.  When the electron beam propagates into the rarefied inner heliosphere,
provided there are enough electrons, wave-particle interaction becomes the energetically dominant effect.
Specifically for our simulations the energy range $6~\rm{keV} \leq E < 40~\rm{keV}$ lost a large percentage of energy
due to resonant interaction with Langmuir waves and the subsequent Langmuir wave interaction with the inhomogeneous background plasma.
The loss of energy caused a spectral flattening of the initial power-law electron distribution
which resulted in a broken power-law (with break around 40 keV) being a very good fit to the data.
We have shown how the broken power-law fit parameters vary as a function of distance,
finding the best fit to the data around 1~AU and the worst fit around 10-30 R$_\odot$.
Finally, we have shown that modelling the turbulent fluctuations in the inner heliosphere is essential
to obtain both an electron distribution and a level of Langmuir waves that is consistent with observations
near the Earth.

It should be stressed that wave-particle interaction not occurring above 40~keV is entirely dependent upon the initial beam
characteristics. Increasing the beam density or decreasing the injected spectral index, characteristic time or size will
all result in a higher number density of electrons that can become unstable to wave-particle interactions
and can consequently have a higher break energy.  Although the parameters we have chosen do reflect typical flare parameters (see Section \ref{ref:Sec3})
we could have run the simulations in a slightly different area of parameter space which would have produced higher or lower break energies.
Other ways to increase the break energy is to inject the electron beam at lower background electron densities
or reduce the expansion factor of the magnetic field.

We showed how the electron beam had to travel around 2 R$_\odot$ before it would become unstable
to the generation of Langmuir waves.  Previous work \cite{Reid_etal2011} showed that the instability distance of an electron beam
was heavily dependent upon the injected spectral index and the longitudinal extent of the acceleration site.
However, the modelled injection of the electron beam was instantaneous in time.  In this paper, we have considered electron beam injection
which is time dependent, with a Gaussian profile and characteristic times, described by Equation (\ref{eqn:time_inj}).
Using a similar analysis to \inlinecite{Reid_etal2011}, we can analyse how a temporal injection alters the instability
distance of an electron beam.  If we assume an injected distribution of the form
\begin{equation}\label{eqn:dist_2}
f(v,r,t)=g_0(v)\exp(-|r|/d)(\pi\tau)^{-\frac{1}{2}}\exp(-|t-t_0|/\tau).
\end{equation}
where both the spatial and temporal form of the electrons is an exponential distribution function.
We obtain an instability criteria of the form
\begin{equation}
h_{typeIII} = (d + v\tau)2\delta + h_{acc}
\end{equation}
where $h_{acc}$ is the injection height from the photosphere and $h_{typeIII}$ is the height that a large level
of Langmuir waves are induced and type III emission is observed.  The instability distance is now dependent
upon the spectral index $\delta$ (2$\delta$ is the spectral index in velocity space) and the highest of the terms in $d+v\tau$.  Assuming an injection site of $10^9$~cm, spectral index $\delta=4$, a characteristic velocity of $5\times10^9~\rm{cm~s}^{-1}$ and a injection time of 10 seconds we find that Langmuir wave
generation will not happen for around $5.5~R_\odot$.  The exact distribution function used in the simulations
varies slightly from Equation (\ref{eqn:dist_2}), but we find Langmuir wave generation occurring around $2.5~R_\odot$
when we consider quasilinear interaction only.  When we consider the collisional term,
we find Langmuir wave growth develops much quicker.  The collisions cause low velocity electrons
near $v_{Te}$ to lose energy, creating a positive gradient in velocity space earlier
than time-of-flight would otherwise produce \cite{HannahKontar2011}.

Finally we note that the spectrum evolves quicker at the distances closer to the Sun, so that future
inner heliosphere missions can diagnose the evolution of energetic electrons, better restricting the possible
parameters of the plasma and energetic electrons. For the parameters adopted in the model,
the simulations predict weakly changing spectral indices and break energy after around $r\sim 20 R_\odot$,
and the decrease of the spectral index below the break at the distances $r<20 R_\odot$. The spectral
break, well pronounced at 1~AU, is less evident at the distances $r<20 R_\odot$, so the peak flux spectrum
will be closer to the electron flux spectrum inferred from hard X-ray observations.

\begin{acks}

 This work is partially supported by a STFC rolling grant (EPK).  The European Commission is acknowledged
 for funding from the HESPE Network (FP7-SPACE-2010-263086) (HASR, EPK) and the SOLAIRE Network (MTRN-CT-2006-035484) (HASR). The overall effort has greatly benefited
 from support by a grant from the Franco-British Alliance Research Programme.  We also thank the referee for useful comments and suggestions.
\end{acks}


\bibliographystyle{spr-mp-sola-cnd}

\bibliography{inner_helio_dec17}

\end{article}

\end{document}